\documentclass[12pt]{article}
\pdfoutput=1

\usepackage{amsmath}
\usepackage{amsfonts}
\usepackage{amscd}
\usepackage{amsthm}
\usepackage{setspace}

\usepackage{graphicx}
\usepackage{authblk}
\usepackage{caption}
\usepackage{ytableau}
\usepackage{mathtools}

\def\Z{\mathbb{Z}}

\def\C{\mathbb{C}}
\def\P{\mathbb{P}}

\def\til{\tilde}

\begin{document}

\begin{titlepage}

\begin{flushright}
KEK-TH 2024
\end{flushright}

\vskip 1cm

\begin{center}

{\large Enhancements in F-theory models on moduli spaces of K3 surfaces with $ADE$ rank 17}

\vskip 1.2cm

Yusuke Kimura$^1$ and Shun'ya Mizoguchi$^{1,2}$
\vskip 0.4cm
{\it $^1$KEK Theory Center, Institute of Particle and Nuclear Studies, KEK, \\ 1-1 Oho, Tsukuba, Ibaraki 305-0801, Japan \\
$^2$SOKENDAI (The Graduate University for Advanced Studies), \\ 1-1 Oho, Tsukuba, Ibaraki 305-0801, Japan}
\vskip 0.4cm
E-mail: kimurayu@post.kek.jp, mizoguch@post.kek.jp 

\vskip 1.5cm
\abstract{We study the moduli of elliptic K3 surfaces with a section with the $ADE$ rank 17. While the Picard number of a generic K3 surface in such moduli space is 19, the Picard number is enhanced to 20 at special points in the moduli. K3 surfaces become attractive K3 surfaces at these points. Either of the following two situations occurs at such special points: i) the Mordell-Weil rank of an elliptic K3 surface is enhanced, or ii) the gauge symmetry is enhanced. The first case i) is related to the appearance of a $U(1)$ gauge symmetry. In this note, we construct the moduli of K3 surfaces with $ADE$ types $E_7 D_{10}$ and $A_{17}$. We determine some of the special points at which K3 surfaces become attractive in the moduli of K3 surfaces with $ADE$ types $E_7 D_{10}$ and $A_{17}$. We investigate the gauge symmetries in F-theory compactifications on attractive K3 surfaces which correspond to such special points in the moduli times a K3 surface. $U(1)$ gauge symmetry arises in some F-theory compactifications.}  

\end{center}
\end{titlepage}

\section{Introduction}
It is now well recognized that F-theory \cite{Vaf, MV1, MV2} offers a useful framework for particle physics model building in string theory. On one hand, F-theory can support 
not only $SU(N)$ or $SO(2N)$ supersymmetric gauge theories achieved in the pure D-brane+orientifold system but an exceptional-group gauge theory, 
which is made possible by identifying all the $(p,q)$ 7-branes 
up to the $SL(2,{\Z})$ monodromy. 
This is particularly relevant for the implementation of GUT 
because it then allows matter in spinor representations, in which 
one generation of quarks and leptons observed in nature perfectly fit. 
On the other hand, another advantage of F-theory is that it is basically 
formulated in type IIB string theory, where the scenarios of moduli 
stabilization and inflation have been well studied, although concrete 
implementations of these in F-theory still  remain a challenge.

\par The rather recent development in F-theory model building has been mainly 
based on local models \cite{DWmodel, BHV, BHV2, DWGUT}, in which the decoupling limit of gravity 
is taken. However, given the results of the recent Large Hadron Collider (LHC) experiment and the 
Planck observation, it is now essential to consider {\em global} models to 
deal with the issues of particle physics beyond the standard model and 
the early universe.  In global models of F-theory, 
compactifications on spaces that admit a K3 fibration are often considered, in which a K3 surface as a fiber of the compactification space is assumed to allow (at least locally) an elliptic fibration. If stable degeneration is available, the sections of the two rational elliptic surfaces encode information on spectral covers of gauge bundles of the dual $E_8\times E_8$ heterotic string theory \cite{FMW}. 
The Picard number of a rational elliptic surface is always 10, and the rank of the singularities and the rank of the Mordell--Weil (MW) group are complementary within $E_8$. The structure of the Mordell--Weil groups of the rational elliptic surfaces was classified by Oguiso-Shioda \cite{OguisoShioda}. 
On the other hand, the Picard numbers of K3 surfaces vary, and such a complementarity relation between the singularities and the MW group does not hold for K3 surfaces.

\par In this note, we study the moduli of K3 surfaces with the $ADE$ rank 17. The N\'eron--Severi lattice of an elliptic surface with a global section is generated by sections, a smooth elliptic fiber and components of the singular fibers \cite{Shiodamodular}. The Picard number is the rank of the N\'eron--Severi lattice, therefore, the following formula holds for an elliptic surface $S$ which admits a section, which is known as the Shioda--Tate formula \cite{Shiodamodular, Tate1, Tate2}:
\begin{equation}
\rho(S)=2+\mbox{rk $ADE$}+ \mbox{rk MW}.
\end{equation}
A generic K3 surface with the $ADE$ rank 17 has the Picard number 19, and the family of K3 surfaces with $ADE$ rank 17 with a fixed $ADE$ type constitutes a 1 dimensional moduli. The Picard number is enhanced to 20 at special points in this moduli, and K3 surfaces become attractive K3 surfaces \footnote{We follow the convention of the term in \cite{M} to refer to K3 surfaces with the Picard number 20 as attractive K3 surfaces. It is standard to refer to a K3 surface with the Picard number 20 as a singular K3 surface in mathematics.} at these points. Either of the following two situations occurs at such special points in the moduli: \\
i)the $ADE$ type is unchanged, and the Mordell--Weil rank is enhanced to 1. \\
ii)the $ADE$ rank is enhanced to 18. \\
A $U(1)$ gauge symmetry arises in F-theory compactifications in the first case, i), and the non-Abelian gauge symmetry in F-theory compactifications is enhanced in the second case, ii). 
\par We particularly construct the moduli of K3 surfaces with a section with the $ADE$ types $E_7 D_{10}$ and $A_{17}$ in this study \footnote{Some relevance of the $E_7$ point 
in particle physics model building was emphasized in \cite{Mizoguchi}.}. We determine some of the special points in these moduli at which K3 surfaces become attractive. Furthermore, we discuss the structures of the gauge groups that arise in F-theory compactifications on these attractive K3 surfaces times a K3 surface. 

\par The orthogonal complement of the N\'eron--Severi lattice $NS(S)$ in the K3 lattice $H^2(S, \Z)$ is called the transcendental lattice, $T(S)$. The transcendental lattice $T(S)$ specifies the complex structure of an attractive K3 surface \cite{PS-S, SI}. The transcendental lattice of an attractive K3 surface is a positive-definite, even lattice, and the intersection matrix of the transcendental lattice of an attractive K3 surface has the size 2 $\times$ 2. A review of the correspondence between the complex structures of the attractive K3 surfaces and the transcendental lattices can be found, e.g., in \cite{BKW, K}. 
\par The following two attractive K3 surfaces, among others, appear at the special points in the moduli of K3 surfaces with the $ADE$ types $E_7 D_{10}$ and $A_{17}$: the attractive K3 surfaces, whose transcendental lattices have the intersection matrices 
\begin{equation}
\begin{pmatrix}
2 & 1 \\
1 & 2 \\
\end{pmatrix} 
\end{equation}
and 
\begin{equation}
\begin{pmatrix}
2 & 0 \\
0 & 2 \\
\end{pmatrix}. 
\end{equation}
The attractive K3 surfaces whose transcendental lattices have the intersection matrices $\begin{pmatrix}
2 & 1 \\
1 & 2 \\
\end{pmatrix}$ and $\begin{pmatrix}
2 & 0 \\
0 & 2 \\
\end{pmatrix}$ are often denoted as $X_3$ and $X_4$, respectively, in the literature; we denote these attractive K3 surfaces as $X_3$ and $X_4$ in this note. The attractive K3 surfaces $X_3$ and $X_4$ are also referred to as the attractive K3 surface with the discriminant 3 and the attractive K3 surface with the discriminant 4, respectively. 
\par Using a method, known as the Kneser-Nishiyama method, the $ADE$ types and the MW groups of the elliptic fibrations with a section of several K3 surfaces, including the attractive K3 surfaces $X_3$ and $X_4$, are classified in \cite{Nish}. 
\par An elliptic K3 surface with a fixed complex structure generally has several different elliptic fibrations. Furthermore, an elliptic K3 surface with a fixed complex structure in general has genus-one fibrations without a section \footnote{F-theory compactifications on genus-one fibrations lacking a global section are studied, for example, in \cite{BM, MTsection, AGGK, KMOPR, GGK, MPTW, MPTW2, BGKintfiber, CDKPP, LMTW, K, K2, KCY4, Kdisc}.}, as well as elliptic fibrations with a global section  \footnote{F-theory models with a section have been discussed, for example, in \cite{MorrisonPark, MPW, BGK, BMPWsection, CKP, LSW}.}. 
\par However, it is known that the attractive K3 surfaces $X_3$ and $X_4$ admit only elliptic fibrations that have a global section. The Weierstrass equations of the elliptic fibrations of the attractive K3 surface $X_3$ are discussed, for example, in \cite{Shioda2008, Sch2008, Ut}. F-theory compactifications on elliptic fibrations of the K3 surfaces $X_3$ and $X_4$ times a K3 surface are discussed in \cite{K, K2, KRES}.
\par In F-theory compactifications, the types of the singular fibers of the compactification space correspond to the non-Abelian gauge groups that form on the 7-branes \cite{MV2, BIKMSV}. We list the types of the singular fibers and the corresponding singularity types of the compactification space in Table \ref{tabfibersingularitycorresp in 2} below. The rank of the Mordell--Weil group gives the number of $U(1)$ gauge fields. 

\begingroup
\renewcommand{\arraystretch}{1.1}
\begin{table}[htb]
\begin{center}
  \begin{tabular}{|c|c|} \hline
Fiber type & $\begin{array}{c}
\mbox{Type of} \\
\mbox{singularity}
\end{array}$ \\ \hline
$I_1$ & none. \\
$II$ & none. \\
$I_N$ ($N\ge 2$) & $A_{N-1}$ \\
$I^*_M$ ($M\ge 0$) & $D_{4+M}$ \\
$III$ & $A_1$ \\
$IV$ & $A_2$ \\
$IV^*$ & $E_6$ \\
$III^*$ & $E_7$ \\
$II^*$ & $E_8$ \\ \hline   
\end{tabular}
\caption{Types of singular fibers and the corresponding types of singularities.}
\label{tabfibersingularitycorresp in 2}
\end{center}
\end{table}  
\endgroup 

\noindent See \cite{Kod1, Kod2} for classification of the types of the singular fibers of elliptic surfaces. Methods to determine the singular fibers of elliptic surfaces are discussed in \cite{Ner, Tate}.

\par The outline of this note is as follows: we review the elliptic fibrations of the attractive K3 surfaces $X_3$ and $X_4$ in section \ref{sec2}. Applying a result in \cite{TataMuk} and analyzing the structure of the Picard lattice \cite{Bel, Keum}, it is shown that every elliptic fibration of the attractive K3 surfaces $X_3$ and $X_4$ has a global section \cite{Keum}. This will be reviewed in section \ref{subsec2.3}. We construct the moduli of K3 surfaces with a section with the $ADE$ types $E_7 D_{10}$ and $A_{17}$ in section \ref{sec 3}. In constructing the moduli of K3 surfaces with the $ADE$ type $E_7 D_{10}$, we use the anomaly cancellation condition to determine the configurations of the singular fibers. The method to obtain the Weierstrass equation by gluing a pair of the identical rational elliptic surfaces is discussed in \cite{KRES}. Applying this method, we obtain the moduli of K3 surfaces with the $ADE$ type $A_{17}$. The Weierstrass equations with parameters describe the moduli of K3 surfaces with the $ADE$ types $E_7 D_{10}$ and $A_{17}$. When the parameters take special values, K3 surfaces become attractive in the moduli. We discuss some of these special points in section \ref{sec 3}. Attractive K3 surfaces $X_3$ and $X_4$ appear at special enhanced points in these moduli. In section \ref{sec 4}, we discuss the gauge groups in F-theory compactifications. We confirm that the anomaly cancellation condition is satisfied when F-theory is compactified on the direct products of K3 surfaces. We also confirm that the tadpole can be cancelled with flux at some special points in the moduli at which K3 surfaces become attractive. We state the concluding remarks in section \ref{sec 5}.

\section{Attractive K3 surfaces $X_3$ and $X_4$}
\label{sec2} 

We discuss the Weierstrass equations of elliptic fibrations of the attractive K3 surfaces $X_3$ and $X_4$.  

\subsection{Weierstrass forms of elliptic fibrations of attractive K3 surface $X_3$}
\label{subsec2.1}
In \cite{Nish}, it was shown that the attractive K3 surface $X_3$ has 6 elliptic fibrations with a global section, and the $ADE$ types and Mordell--Weil groups of these fibrations were obtained. We list the results of these 6 elliptic fibrations obtained in \cite{Nish} in Table \ref{tabfibrationX3} below. We discuss the Weierstrass forms of the 6 elliptic fibrations with a section of the K3 surface $X_3$. 

\begingroup
\renewcommand{\arraystretch}{1.1}
\begin{table}[htb]
\centering
  \begin{tabular}{|c|c|c|} \hline
Fibration & $ADE$ type & MW group \\ \hline
No.1 & $E^2_8 A_2$ & 0 \\ 
No.2 &  $D_{16} A_2$ & $\Z_2$ \\
No.3 & $E_7 D_{10}$ & $\Z\oplus \Z_2$ \\
No.4 & $A_{17}$ & $\Z\oplus \Z_3$ \\ 
No.5 & $E^3_6$ & $\Z_3$ \\
No.6 & $D_7 A_{11}$ & $\Z_4$ \\ \hline
\end{tabular}
\caption{\label{tabfibrationX3}Elliptic fibrations of the K3 surface $X_3$, their $ADE$ types and the Mordell--Weil groups.}
\end{table}
\endgroup

\vspace{1cm}

Fibration No.1 has the following Weierstrass form \cite{ShiodaKummer, Shioda2008} :
\begin{equation}
\label{no1 fibration Weierstrass in 2.1}
y^2=x^3+t^5 \, s^5 \, (t-s)^2.
\end{equation}
We have used the notation $[t:s]$ to denote the homogeneous coordinate on the base $\P^1$.  
\par The $ADE$ type of fibration No.1 is $E^2_8 A_2$. We see from Table 2 in \cite{SZ} that the K3 surface $X_3$ is a unique extremal elliptic K3 surface \footnote{An extremal K3 surface is an elliptically fibered attractive K3 surface, which admits a global section, with the Mordell--Weil rank 0.} with $ADE$ type $E^2_8 A_2$. We find from equation (\ref{no1 fibration Weierstrass in 2.1}) that the fibration No.1 has 2 type $II^*$ fibers, at $[t:s]=[0:1]$ and $[1:0]$, and 1 type $IV$ fiber at $[t:s]=[1:1]$. We list the vanishing orders of zero of the coefficients, $f,g$, and the discriminant $\Delta$, of the Weierstrass form $y^2=x^3+f\, x+g$ for the types of the singular fibers in Table \ref{tabWeierstrassorder}.

\begingroup
\renewcommand{\arraystretch}{1.5}
\begin{table}[htb]
\begin{center}
  \begin{tabular}{|c|c|c|c|} \hline
Fiber type & Ord($f$) & Ord($g$) & Ord($\Delta$) \\ \hline
$I_0 $ & $\ge 0$ & $\ge 0$ & 0 \\ \hline
$I_n $  ($n\ge 1$) & 0 & 0 & $n$ \\ \hline
$II $ & $\ge 1$ & 1 & 2 \\ \hline
$III $ & 1 & $\ge 2$ & 3 \\ \hline
$IV $ & $\ge 2$ & 2 & 4 \\ \hline
$I_0^*$ & $\ge 2$ & 3 & 6 \\ \cline{2-3}
 & 2 & $\ge 3$ &  \\ \hline
$I_m^*$  ($m \ge 1$) & 2 & 3 & $6+m$ \\ \hline
$IV^*$ & $\ge 3$ & 4 & 8 \\ \hline
$III^*$ & 3 & $\ge 5$ & 9 \\ \hline
$II^*$ & $\ge 4$ & 5 & 10 \\ \hline   
\end{tabular}
\caption{\label{tabWeierstrassorder}The vanishing orders of coefficients $f,g$ of the Weierstrass form $y^2=x^3+f\, x+g$, and the vanishing order of the discriminant $\Delta$, for types of the singular fibers.}
\end{center}
\end{table}  
\endgroup 

\vspace{5mm}

The Weierstrass form of the fibration No.5 is given as follows:
\begin{equation}
\label{no5 fibration Weierstrass}
y^2=x^3+(t-\alpha_1 \, s)^4 (t-\alpha_2 \, s)^4 (t-\alpha_3 \, s)^4.
\end{equation}
In equation (\ref{no5 fibration Weierstrass}), $\alpha_i$, $i=1,2,3$, are constants, and they are mutually distinct. From the vanishing orders of the coefficient, we find that the Weierstrass form (\ref{no5 fibration Weierstrass}) has three singular fibers of type $IV^*$. The fibration No.5 has the $ADE$ type $E^3_6$, and, as discussed in \cite{KCY4}, it can be deduced from Table 2 in \cite{SZ} that the K3 surface $X_3$ is a unique extremal K3 surface with the $ADE$ type $E^3_6$. 

\vspace{1cm}

Smooth fibers of the fibrations No.1 and No.5 admit transformations into the Fermat curve. This can be seen as follows: in homogeneous coordinates, fibers of the fibrations No.1 and No.5 are described by equations of the following form:
\begin{equation}
\label{fiber no1no5}
y^2z=x^3+\beta\, z^3.
\end{equation}
After some rescaling, equation (\ref{fiber no1no5}) may be rewritten as 
\begin{equation}
\label{trans no1no5}
x^3+z(z^2+y^2)=0.
\end{equation}
Using the following transformation
\begin{eqnarray}
z & = & \frac{1}{\sqrt[3]{4}}(\til{z}+\til{y}) \\ \nonumber
y & = & \frac{\sqrt{3}}{\sqrt[3]{4}} (\til{y}-\til{z}), 
\end{eqnarray}
equation (\ref{trans no1no5}) transforms into the Fermat curve:
\begin{equation}
x^3+\til{y}^3+\til{z}^3=0.
\end{equation}
Fermat curves have j-invariant 0, therefore, smooth fibers of the fibrations No.1 and No.5 have the constant j-invariant 0 over the base $\P^1$. Thus, it follows that the singular fibers of both the fibrations No.1 and No.5 have j-invariant 0. 
\par Each singular fiber type has a specific j-invariant. We list the correspondence of the fiber types and j-invariants \cite{Kod1} in Table \ref{tabjinvariant}. 
\begingroup
\renewcommand{\arraystretch}{1.1}
\begin{table}[htb]
\centering
  \begin{tabular}{|c|c|r|c|} \hline
Fiber Type & j-invariant & Monodromy  & order of Monodromy \\ \hline
$I^*_0$ & finite & $-\begin{pmatrix}
1 & 0 \\
0 & 1 \\
\end{pmatrix}$ & 2 \\ \hline
$I_b$ & $\infty$ & $\begin{pmatrix}
1 & b \\
0 & 1 \\
\end{pmatrix}$ & infinite \\
$I^*_b$ & $\infty$ & $-\begin{pmatrix}
1 & b \\
0 & 1 \\
\end{pmatrix}$ & infinite \\ \hline
$II$ & 0 & $\begin{pmatrix}
1 & 1 \\
-1 & 0 \\
\end{pmatrix}$ & 6 \\
$II^*$ & 0 & $\begin{pmatrix}
0 & -1 \\
1 & 1 \\
\end{pmatrix}$ & 6 \\ \hline
$III$ & 1728 & $\begin{pmatrix}
0 & 1 \\
-1 & 0 \\
\end{pmatrix}$ & 4 \\
$III^*$ & 1728 & $\begin{pmatrix}
0 & -1 \\
1 & 0 \\
\end{pmatrix}$ & 4 \\ \hline
$IV$ & 0 & $\begin{pmatrix}
0 & 1 \\
-1 & -1 \\
\end{pmatrix}$ & 3 \\
$IV^*$ & 0 & $\begin{pmatrix}
-1 & -1 \\
1 & 0 \\
\end{pmatrix}$ & 3 \\ \hline
\end{tabular}
\caption{Fiber types and the corresponding j-invariants. ``Finite'' for type $I^*_0$ fiber means that j-ivariant of $I^*_0$ fiber can take any finite value in $\C$. Monodromies around the singular fibers \cite{Kod1}, and their orders are also included.}
\label{tabjinvariant}
\end{table}
\endgroup 
We deduce from Table \ref{tabjinvariant} that the fiber types that have j-invariant 0 are only $II$, $IV$, $IV^*$, $I^*_0$ and $II^*$ (fiber type $I^*_0$ may take j-invariant 0; the j-invariant of this fiber type takes any finite value in $\C$, depending on the situation.) The types of the singular fibers of the fibration No.1 are $II^*$ and $IV$, and the fibration No.5 has the singular fibers of type $IV^*$. We confirm that these are in agreement with the constraints imposed by the j-invariant of the singular fibers \footnote{When elliptic fibers possess complex multiplications of orders 3 and 4, throughout the base, elliptic fibers have constant j-invariants 0 and 1728, respectively, over the base. The possible non-Abelian gauge groups that can form on the 7-branes in F-theory compactifications are strongly constrained by these symmetries \cite{K, K2, KCY4}. Fermat curve possesses complex multiplication of order 3.}.

\vspace{1cm}

The fibration No.2 has the $ADE$ type $D_{16} A_2$. According to Table 2 in \cite{SZ}, an extremal K3 surface with the $ADE$ type $D_{16} A_2$ is uniquely determined to be the K3 surface $X_3$. Therefore, it is enough to construct the Weierstrass form that has the $ADE$ type $D_{16} A_2$ to determine the Weierstrass equation of the fibration No.2. 
\par As discussed in \cite{Sch2007}, the triple cover of the extremal rational elliptic surface \footnote{A rational elliptic surface is said to be extremal when the $ADE$ type has the rank 8. The extremal rational elliptic surfaces were classified in \cite{MP}. See, for example, \cite{KRES} for a review of the extremal rational elliptic surfaces.}, $X_{[4^*, \hspace{1mm} 1, 1]}$, is an attractive K3 surface with 1 $I^*_{12}$ fiber, 1 $I_3$ fiber, and 3 $I_1$ fibers. (See No. 216 in Table 2 in \cite{Sch2007}.) The extremal rational elliptic surface $X_{[4^*, \hspace{1mm} 1, 1]}$ is the extremal rational elliptic surface with three singular fibers, of types $I^*_4$, $I_1$, and $I_1$. An extremal rational elliptic surface with such configuration of singular fibers is unique \cite{MP}. We follow the notation used in \cite{KRES} to denote by $X_{[4^*, \hspace{1mm} 1, 1]}$ the extremal rational elliptic surface with three singular fibers, of types $I^*_4$, $I_1$, and $I_1$, in this study. The surface $X_{[4^*, \hspace{1mm} 1, 1]}$ is given by the following Weierstrass form \cite{MP}:
\begin{equation}
\label{411 in 2.1}
y^2=x^3-3t^2(s^2-3t^2)\, x+t^3s\, (2s^2-9t^2).
\end{equation}
The following linear transformation of the Weierstrass form (\ref{411 in 2.1}) was considered in \cite{Sch2007}:
\begin{equation}
\label{lineartransf 411}
(t,\, s) \rightarrow (t, \, 4s-2t).
\end{equation}
Under the linear transformation (\ref{lineartransf 411}), equation (\ref{411 in 2.1}) transforms into another Weierstrass form as:
\begin{equation}
\label{another Weierstrass 411 in 2.1}
y^2=x^3-3t^2\, (t^2-16ts+16s^2)\, x+2t^3\, (t-2s) (t^2+32ts-32s^2).
\end{equation}
We consider the following triple cover of the Weierstrass form (\ref{another Weierstrass 411 in 2.1}) of the extremal rational elliptic surface $X_{[4^*, \hspace{1mm} 1, 1]}$ :
\begin{equation}
\label{base change 411 in 2.1}
(t,\, s) \rightarrow (t^3, \, s^3).
\end{equation}
The resulting K3 surface is given by the following Weierstrass form \footnote{When we take the base change (\ref{base change 411 in 2.1}), we obtain the following Weierstrass form:
\begin{equation}
\label{nonminimal Weierstrass no2 in 2.1}
y^2=x^3-3t^6(t^6-16t^3s^3+16s^6)\, x+2t^9(t^3-2s^3)(t^6+32t^3s^3-32s^6).
\end{equation}
The Weierstrass equation (\ref{nonminimal Weierstrass no2 in 2.1}) is not minimal; we consider the following minimalizing process:
\begin{equation}
x \rightarrow t^2 \, x, \hspace{1cm} y\rightarrow t^3\, y.
\end{equation}
Consequently, we obtain the minimal Weierstrass equation (\ref{no2 Weierstrass in 2.1}). Minimalizing process is discussed in \cite{MP2} in the context of F-theory.}
 :
\begin{equation}
\label{no2 Weierstrass in 2.1}
y^2=x^3-3t^2(t^6-16t^3s^3+16s^6)\, x+2t^3(t^3-2s^3)(t^6+32t^3s^3-32s^6),
\end{equation}
with the discriminant 
\begin{equation}
\label{disc no2}
\Delta \sim t^{18}s^3 \, (t^3-s^3).
\end{equation}
From the equations (\ref{no2 Weierstrass in 2.1}) and (\ref{disc no2}), we find that the K3 surface (\ref{no2 Weierstrass in 2.1}) has 1 type $I_{12}^*$ fiber at $[t:s]=[0:1]$, 1 type $I_3$ fiber at $[t:s]=[1:0]$, and 3 type $I_1$ fibers at the roots of $t^3-s^3=0$. 
\par Therefore, the K3 surface (\ref{no2 Weierstrass in 2.1}) has the $ADE$ type $D_{16} A_2$. Thus, we conclude that the Weierstrass equation (\ref{no2 Weierstrass in 2.1}) describes the fibration No.2.

\vspace{1cm}

\par We discuss the Weierstrass forms of the fibrations No.3 and No.4 separately, in sections \ref{subsec3.1.1} and \ref{subsec3.2.1}, respectively. They correspond to special points in the moduli of K3 surfaces with the $ADE$ types $E_7 D_{10}$ and $A_{17}$, respectively, at which the Mordell--Weil rank is enhanced to 1. 

\vspace{1cm}

The general Weierstrass form of the fibration No.6 can be found in \cite{Sch2008}; the general Weierstrass form is given by:
\begin{equation}
\label{no6 general Weierstrass in 2.1}
y^2+t^2\, xy=x^3+2t\, x^2+t^2\, x.
\end{equation}
The general Weierstrass equation (\ref{no6 general Weierstrass in 2.1}) transforms into the following Weierstrass form:
\begin{equation}
\label{no6 Weierstrass in 2.1}
y^2=x^3+t^2\,[ \, s^6-\frac{1}{48}(t^3+8s^3)^2 \, ]\, x+\frac{t^3}{27}(\frac{t^3}{4}+2s^3)(\frac{t^6}{8}+2t^3\, s^3-s^6).
\end{equation}
The discriminant of the fibration No.6 (\ref{no6 Weierstrass in 2.1}) is 
\begin{equation}
\Delta \sim t^9(t^3+16s^3)\, s^{12}.
\end{equation}
The $I^*_3$ fiber is at $[t:s]=[0:1]$, the $I_{12}$ fiber is at $[t:s]=[1:0]$, and three $I_1$ fibers are at the roots of $t^3+16s^3=0$ \cite{Sch2008}. 

\vspace{5mm}

\par The fibrations No.1 and  No.5 are discussed in \cite{KRES} in the context of the stable degenerations of K3 surfaces into pairs of rational elliptic surfaces. We construct the moduli of K3 surfaces with the $ADE$ type $A_{17}$ by gluing two identical rational elliptic surfaces, and taking a special limit, in section \ref{subsec3.2.1}. We find that the fibration No.4 in Table \ref{tabfibrationX3} corresponds to a special point in this moduli with the $ADE$ type $A_{17}$ at which the Mordell--Weil rank is enhanced to 1.

\subsection{Weierstrass forms of elliptic fibrations of attractive K3 surface $X_4$}
The $ADE$ types and the Mordell--Weil groups of the elliptic fibrations with a section of the attractive K3 surface with the discriminant 4, $X_4$, were classified in \cite{Nish}. We list the results deduced in \cite{Nish} in Table \ref{tabfibrationX4} below. 

\begingroup
\renewcommand{\arraystretch}{1.1}
\begin{table}[htb]
\centering
  \begin{tabular}{|c|c|c|} \hline
Fibration & $ADE$ type & MW group \\ \hline
No.1 & $E^2_8 A^2_1$ & 0 \\ 
No.2 &  $E_8 D_{10}$ & 0 \\
No.3 & $D_{16} A^2_1$ & $\Z_2$ \\
No.4 & $E^2_7 D_4$ & $\Z_2$ \\ 
No.5 & $E_7 D_{10} A_1$ & $\Z_2$ \\
No.6 & $A_{17} A_1$ & $\Z_3$ \\ 
No.7 & $D_{18}$ & 0 \\ 
No.8 &  $D_{12} D_6$ & $\Z_2$ \\
No.9 & $D^2_8 A^2_1$ & $\Z_2 \oplus \Z_2$ \\
No.10 & $A_{15} A_3$ & $\Z_4$ \\ 
No.11 & $E_6 A_{11}$ & $\Z\oplus \Z_3$ \\
No.12 & $D^3_6$ & $\Z_2\oplus\Z_2$ \\ 
No.13 & $A^2_9$ & $\Z_5$ \\ \hline
\end{tabular}
\caption{\label{tabfibrationX4}Elliptic fibrations of the K3 surface $X_4$, their $ADE$ types and the Mordell--Weil groups.}
\end{table}
\endgroup

We discuss some of the elliptic fibrations with a global section of the attractive K3 surface $X_4$. 

\vspace{5mm}

\par The fibration No.1 in Table \ref{tabfibrationX4} has the following Weierstrass form \cite{ShiodaKummer, Shioda2008}:
\begin{equation}
\label{Weierstrass no1 X_4}
y^2=x^3-3\, t^4s^4\,x +t^5s^5\, (t^2+s^2).
\end{equation}
The discriminant of the Weierstrass form (\ref{Weierstrass no1 X_4}) is 
\begin{equation}
\label{disc no1 X_4}
\Delta \sim t^{10}s^{10}(t+s)^2(t-s)^2.
\end{equation}
It can be seen from (\ref{Weierstrass no1 X_4}) and (\ref{disc no1 X_4}) that there are 2 type $II^*$ fibers, at $[t:s]=[1:0]$ and at $[0:1]$. There are 2 type $I_2$ fibers, at $[t:s]=[1:1]$ and at $[1:-1]$. 

\vspace{5mm}

\par The fibration No.4 has the following Weierstrass form:
\begin{equation}
\label{Weierstrass no4 X_4}
y^2=x^3+ (t-s)^2\, t^3s^3\, x.
\end{equation}
The discriminant of the Weierstrass equation (\ref{Weierstrass no4 X_4}) \footnote{The smooth elliptic fibers of the Weierstrass equation (\ref{Weierstrass no4 X_4}) have complex multiplication of order 4.} is 
\begin{equation}
\label{disc no4 X_4}
\Delta \sim t^9s^9\, (t-s)^6.
\end{equation}
We find from (\ref{Weierstrass no4 X_4}) and (\ref{disc no4 X_4}) that there are 2 type $III^*$ fibers, at $[t:s]=[1:0]$ and at $[t:s]=[0:1]$. There is 1 type $I^*_0$ fiber at $[t:s]=[1:1]$.  

\vspace{5mm} 

\par The fibration No.9 is given by the following Weierstrass equation \cite{KRES} :
\begin{equation}
y^2=x^3-3\, t^2s^2\, (t^4+s^4-t^2s^2)\, x+ (t^2+s^2)\, t^3s^3 \, (2t^4-5t^2s^2+2s^4),
\end{equation}
with the discriminant 
\begin{equation}
\Delta \sim t^{10} s^{10}\, (t-s)^2(t+s)^2.
\end{equation}
There are 2 type $I^*_4$ fibers, at $[t:s]=[0:1]$ and $[1:0]$, and there are 2 type $I_2$ fibers, at $[t:s]=[1:1]$ and $[1:-1]$.

\vspace{5mm} 

We will discuss the fibrations No.2, 5 and 6 in section \ref{sec 3} as special points of the moduli of K3 surfaces with the $ADE$ types $E_7D_{10}$ and $A_{17}$, at which the $ADE$ ranks are enhanced. 
\par The fibrations No.1, 4, 9 are discussed in \cite{KRES} in the context of the stable degenerations of K3 surfaces into pairs of rational elliptic surfaces.

\subsection{Absence of genus-one fibration without a section in K3 surfaces $X_3$ and $X_4$}
\label{subsec2.3}
In \cite{Keum}, it is shown that every elliptic fibration of the K3 surfaces $X_3$ and $X_4$ has a global section. The argument goes as follows: If the K3 surface $X_3$ has a genus-one fibration that lacks a global section, we can consider the Jacobian fibration $J(X_3)$ of this genus-one fibration. The Jacobian fibration of a genus-one fibration is an elliptic fibration with a section, the types and locations of the singular fibers of which are identical to those of the singular fibers of the genus-one fibration. 
\par We denote the smallest degree of the multisections that the genus-one fibration of the K3 surface $X_3$ possesses by $n$, with $n>1$. (The multisection of degree $n=1$ is a global section.) Then, the following relationship holds \cite{Keum} between the determinants of the Picard lattice $Pic(X_3)$ of the genus-one fibration, and the Picard lattice $Pic(J(X_3))$ of the Jacobian fibration $J(X_3)$:
\begin{equation}
{\rm det}\, Pic\, (X_3) = n^2 \, {\rm det}\, Pic\, (J(X_3)).
\end{equation}
Since the transcendental lattice is the orthogonal complement of the Picard lattice in the K3 lattice, we have \cite{Nik}
\begin{equation}
{\rm det}\, Pic\, (X_3) = {\rm det}\, T(X_3).
\end{equation}
We have 
\begin{equation}
{\rm det}\, T(X_3) = {\rm det}\, \begin{pmatrix}
2 & 1 \\
1 & 2 \\
\end{pmatrix} =3,
\end{equation}
therefore, we obtain
\begin{equation}
\label{det eqn in 2.4}
n^2 \, {\rm det}\, Pic\, (J(X_3)) = 3.
\end{equation}
Equation (\ref{det eqn in 2.4}) implies that $n^2$ divides 3. However, $n^2$ divides 3 only when $n=1$; $n=1$ means that the genus-one fibration has a global section, which is a contradiction. Therefore, the K3 surface $X_3$ does not have a genus-one fibration without a global section. 
\par An argument similar to that stated above shows that every elliptic fibration of the attractive K3 surface $X_4$ has a global section. This can be seen as follows: The intersection matrix of the transcendental lattice of an attractive K3 surface can be transformed into the following form under the $GL_2(\Z)$ action:
\begin{equation}
\begin{pmatrix}
2a & b \\
b & 2c 
\end{pmatrix},
\end{equation}
where $a,b,c\in \Z$, and $a\ge c\ge b\ge 0$. $a$ and $c$ are positive integers. Thus, the following inequality holds for the discriminant of an attractive K3 surface:
\begin{equation}
{\rm det}\, \begin{pmatrix}
2a & b \\
b & 2c 
\end{pmatrix} = 4ac-b^2 \ge 4c^2-c^2=3c^2\ge 3.
\end{equation}
It follows that the minimum of the discriminants of the attractive K3 surfaces is 3. Suppose that the K3 surface $X_4$ admits a genus-one fibration lacking a global section. We consider the Jacobian fibration $J(X_4)$ of this genus-one fibration. Since the K3 surface $X_4$ has the discriminant 4, we find from the relationship \cite{Keum} between the determinants of the Picard lattice $Pic(X_4)$ of the genus-one fibration, and the Picard lattice $Pic(J(X_4))$ of the Jacobian fibration $J(X_4)$, that the degree of the multisection of this genus-one fibration must be 2. This means that the Jacobian fibration $J(X_4)$ has the discriminant 1, which contradicts the fact that the minimum of the discriminants of the attractive K3 surfaces is 3. 
\par In particular, 6 elliptic fibrations with a section deduced in \cite{Nish} classify the elliptic fibrations of the attractive K3 surface $X_3$. Similarly, 13 elliptic fibrations with a section obtained in \cite{Nish} classify the elliptic fibrations of the K3 surface $X_4$. 
\par Applying an argument similar to the proof stated above, we see that every elliptic fibration of the attractive K3 surface whose transcendental lattice has the intersection matrix $\begin{pmatrix}
4 & 0 \\
0 & 2 \\
\end{pmatrix}$ has a section. Suppose that the attractive K3 surface whose transcendental lattice has the intersection matrix $\begin{pmatrix}
4 & 0 \\
0 & 2 \\
\end{pmatrix}$ has a genus-one fibration without a global section. This attractive K3 surface has the discriminant 8, therefore, the degree of the genus-one fibration without a section must be 2. This means that the discriminant of the Jacobian fibration of the genus-one fibration lacking a section is 2. However, the smallest discriminant of the attractive K3 surfaces is 3, which is a contradiction. Thus, every elliptic fibration of the attractive K3 surface whose transcendental lattice has the intersection matrix $\begin{pmatrix}
4 & 0 \\
0 & 2 \\
\end{pmatrix}$ has a global section.

\section{Moduli of K3 surfaces with $ADE$ types $E_7 D_{10}$ and $A_{17}$}
\label{sec 3}

\subsection{Moduli of K3 surfaces with $ADE$ type $E_7 D_{10}$}
\label{subsec3.1}

\subsubsection{Weierstrass equation of K3 surfaces with $ADE$ type $E_7 D_{10}$}
\label{subsec3.1.1}
We determine the form of the Weierstrass equation of a general K3 surface with $ADE$ type $E_7 D_{10}$. 
\par First, we specify the configuration of singular fibers of a K3 surface with $ADE$ type $E_7 D_{10}$. We consider F-theory compactification on a K3 surface with $ADE$ type $E_7 D_{10}$ times a K3. The form of the discriminant locus in the base space $\P^1\times$ K3 is determined by the anomaly cancellation conditions to be 24 K3 surfaces. For example, see \cite{K}, for discussion of the form of the discriminant locus in F-theory compactification on the direct product of K3 surfaces. This imposes a constraint on the configuration of the singular fibers of K3 surface; the sum of the number of 7-branes associated with the singular fibers must be 24. The number of the 7-branes associated to a singular fiber is given by the Euler number of that fiber. The Euler numbers of the singular fibers are given in \cite{Kod2}. We list the numbers of 7-branes for the singular fiber types in Table \ref{tab7-branesnumber}. 
\begingroup
\renewcommand{\arraystretch}{1.1}
\begin{table}[htb]
\centering
  \begin{tabular}{|c|c|} \hline
$\begin{array}{c}
\mbox{Type of} \\
\mbox{ singular fiber}
\end{array}$ & $\begin{array}{c}
\mbox{\# of 7-branes} \\
\mbox{(Euler number)}
\end{array}$ \\ \hline
$I_n$ & $n$\\
$I_0^*$ &  6\\ 
$I_m^*$ &  $m+$6\\ 
$II $ &  2\\
$III $ & 3\\
$IV $ & 4\\
$IV^*$ & 8 \\ 
$III^*$ & 9 \\
$II^*$ & 10 \\ \hline
\end{tabular}
\caption{\label{tab7-branesnumber}Types of singular fibers and the associated numbers of 7-branes.}
\end{table}
\endgroup
From the constraint that the total number of 7-branes is 24, we deduce that only two configurations of singular fibers are possible for $ADE$ type $E_7 D_{10}$: \\
i) 1 type $III^*$ fiber, 1 type $I^*_6$ fiber, and 3 type $I_1$ fibers \\
ii) 1 type $III^*$ fiber, 1 type $I^*_6$ fiber, 1 type $II$ fiber, and 1 type $I_1$ fiber. \\
\par First, we discuss K3 surfaces with the second configuration, ii), of singular fibers. We may assume that the type $I^*_6$ fiber is at $[t:s]=[0:1]$, the type $III^*$ fiber is at $[t:s]=[1:0]$, and the type $II$ fiber is at $[t:s]=[1:1]$. We find from Table \ref{tabWeierstrassorder} that the coefficients $f,g$ of the Weierstrass form \footnote{Elliptic K3 surfaces are elliptic fibrations over the base $\P^1$, therefore, the Weierstrass equations that we consider in this note are elliptic curves over the function field $\C(u)$, where $u:=t/s$.} $y^2=x^3+f\, x+g$ of configuration ii) must be of the following form: 
\begin{eqnarray}
\label{array congig 2 in 3.1.1}
f & = & t^2 \, s^3 \, (t-s) \, h_2 \\ \nonumber
g & = & t^3 \, s^5 \, (t-s) \, h_3.
\end{eqnarray}
In equation (\ref{array congig 2 in 3.1.1}), we have used $h_2$, $h_3$ to denote homogeneous polynomials in $t,s$ of degrees 2 and 3, respectively. The discriminant is given as follows:
\begin{equation}
\label{disc config 2 in 3.1.1}
\Delta = t^6 \, s^9 \, (t-s)^2 \, [\, 4(t-s)\, h^3_2+27s\, h^2_3 \,].
\end{equation}
To have a type $I^*_6$ fiber at $[t:s]=[0:1]$, $t^6$ must divide the term 
\begin{equation}
4(t-s)\, h^3_2+27s\, h^2_3
\end{equation}
in the discriminant (\ref{disc config 2 in 3.1.1}). The following polynomials $h_2, h_3$ satisfy this condition:
\begin{eqnarray}
\label{solution config 2 in 3.1.1}
h_2 & = & \frac{\alpha}{6}((i\sqrt{3}-3)t^2-6i\sqrt{3}ts+6s^2) \\ \nonumber 
h_3 & = & \frac{\sqrt{\alpha^3}}{18}(-(5\sqrt{3}+3i)t^3+4(2\sqrt{3}-3i)t^2s+2(\sqrt{3}+9i)ts^2-4\sqrt{3}s^3).
\end{eqnarray}
Under the following rescaling
\begin{eqnarray}
\label{rescaling in 3.1.1}
x & \rightarrow & \sqrt{\alpha}\, x \\ \nonumber
y & \rightarrow & \alpha^{3/4}\, y,
\end{eqnarray}
the coefficients $f,g$ of the Weierstrass form $y^2=x^3+f\, x+g$ transform as
\begin{eqnarray}
\label{rescaling coeff in 3.1.1}
f & \rightarrow & \frac{f}{\alpha} \\ \nonumber
g & \rightarrow & \frac{g}{\sqrt{\alpha^3}}.
\end{eqnarray}
Therefore, the complex structures of the elliptic K3 surfaces, the coefficients of the Weierstrass form of which are given by (\ref{array congig 2 in 3.1.1}) where $h_2$ and $h_3$ are given by (\ref{solution config 2 in 3.1.1}), in fact do not depend on the values of $\alpha$ in (\ref{solution config 2 in 3.1.1}); we may fix the parameter $\alpha$ as $\alpha=1$. The moduli of the elliptic K3 surfaces with a section, the singular fibers of which have configuration ii), is 0-dimensional, namely, it is discrete. Particularly, this means that neither enhancement of the singularity type nor enhancement of the Mordell--Weil rank occurs for the moduli of these elliptic K3 surfaces. We do not consider the elliptic K3 surfaces with a section, the singular fibers of which have configuration ii), in this study.
\par We discuss K3 surfaces with $ADE$ type $E_7 D_{10}$ that have the first configuration, i), of the singular fibers: 1 type $III^*$ fiber, 1 type $I^*_6$ fiber, and 3 type $I_1$ fibers. We determine the form of the Weierstrass equation which has this configuration of singular fibers. We may assume that the type $I^*_6$ fiber is at $[t:s]=[0:1]$, and the type $III^*$ fiber is at $[t:s]=[1:0]$. The coefficients $f,g$ of the Weierstrass form $y^2=x^3+f\, x+g$ must be of the following form:
\begin{eqnarray}
f & = & t^2 \, s^3 \, \til{h}_3 \\ \nonumber
g & = & t^3 \, s^5 \, \til{h}_4.
\end{eqnarray}
We have used $\til{h}_3, \til{h}_4$ to denote homogeneous polynomials in $t,s$ of degree 3 and 4, respectively. The discriminant is given by
\begin{equation}
\label{disc config 1 in 3.1.1}
\Delta = t^6 \, s^9 \, (4\til{h}^3_3+27s\, \til{h}^2_4).
\end{equation}
We have a type $I^*_6$ fiber at $[t:s]=[0:1]$; therefore, we impose the condition that $t^6$ divides the term 
\begin{equation}
\label{term in 3.1.1}
4\til{h}^3_3+27s\, \til{h}^2_4
\end{equation}
in the discriminant (\ref{disc config 1 in 3.1.1}).
Solving this condition, we obtain two families of the Weierstrass equations that give the elliptic K3 surfaces with a global section with $ADE$ type $E_7 D_{10}$. 
\par One family of K3 surfaces with $ADE$ type $E_7 D_{10}$ is given by the following Weierstrass form:
\begin{equation}
\label{Weierstrass general in 3.1.1}
\begin{split}
y^2= & x^3+ t^2 \, s^3 \, (\alpha \, t^3 -\frac{1}{12}\beta^2 \, t^2 s+ \beta\, ts^2 -3s^3)\, x \\
& + t^3 \, s^5 \, [\, -\frac{\alpha\beta}{6}t^4 + (\alpha+\frac{\beta^3}{108})\, t^3s-\frac{1}{6}\beta^2\, t^2s^2 +\beta \, ts^3-2s^4 \, ].
\end{split}
\end{equation}
In the Weierstrass equation (\ref{Weierstrass general in 3.1.1}), $(\alpha, \beta)$ are parameters \footnote{The solution to the condition that $t^6$ divides the term (\ref{term in 3.1.1}) has three parameters. Using the rescaling similar to (\ref{rescaling in 3.1.1}) and (\ref{rescaling coeff in 3.1.1}), one of the three parameters can be fixed. Consequently, we obtain the family (\ref{Weierstrass general in 3.1.1}).}. The discriminant of the Weierstrass form (\ref{Weierstrass general in 3.1.1}) is 
\begin{equation}
\label{disc E7D10 general in 3.1.1}
\Delta \sim t^{12}\, s^9 \, (4\alpha\, t^3-\frac{\beta^2}{4} \, t^2s+3\beta\, ts^2-9s^3).
\end{equation}
\par From equations (\ref{Weierstrass general in 3.1.1}) and (\ref{disc E7D10 general in 3.1.1}), we confirm that the type $I^*_6$ fiber is at $[t:s]=[0:1]$, the type $III^*$ fiber is at $[t:s]=[1:0]$, and three type $I_1$ fibers are at the roots of $4\alpha\, t^3-\frac{\beta^2}{4} \, t^2s+3\beta\, ts^2-9s^3$, for generic values of $(\alpha, \beta)$. Thus, we confirm that the $ADE$ type of K3 surface (\ref{Weierstrass general in 3.1.1}) is in fact $E_7 D_{10}$.
\par The Weierstrass form of the other family of elliptic K3 surfaces with a section with the $ADE$ type $E_7 D_{10}$ is given as:
\begin{equation}
\label{Weierstrass the other in 3.1.1}
\begin{split}
y^2 = & x^3 + t^2s^3\, [\, \frac{-1}{144} \, (12\gamma\delta+\delta^3)\, t^3+\gamma \, t^2s +\delta \, ts^2 -3s^3 \, ]\, x \\
& +t^3s^5\, [\, (-\frac{\gamma^2}{12}+\frac{\delta^4}{1728})\, t^4-(\frac{\gamma\delta}{4}+\frac{5}{432}\delta^3)\, t^3s+(\gamma-\frac{\delta^2}{12})\, t^2s^2+\delta \, ts^3-2s^4 \, ].
\end{split}
\end{equation}
$(\gamma, \delta)$ are parameters. 
\par The discriminant of the Weierstrass equation (\ref{Weierstrass the other in 3.1.1}) is given as follows:
\begin{equation}
\label{disc the other in 3.1.1}
\begin{split}
\Delta \sim t^{12}\, s^{9}\, &  [\, -\frac{1}{324}(12\gamma\delta+\delta^3)\, \delta^2\,t^3+\frac{1}{48}(4\gamma+\delta^2)\, (36\gamma+\delta^2)\, t^2s \\
& +\frac{1}{2}\, \delta (4\gamma+\delta^2)\, ts^2-\frac{1}{3}\, (24\gamma+5\delta^2)\, s^3 \,].
\end{split}
\end{equation}
We find from the discriminant (\ref{disc the other in 3.1.1}) that there is 1 type $III^*$ fiber at $[t:s]=[1:0]$, 1 type $I^*_6$ fiber at $[t:s]=[0:1]$, and there are 3 type $I_1$ fibers at the roots of $-\frac{1}{324}(12\gamma\delta+\delta^3)\, \delta^2\,t^3+\frac{1}{48}(4\gamma+\delta^2)\, (36\gamma+\delta^2)\, t^2s+\frac{1}{2}\, \delta (4\gamma+\delta^2)\, ts^2-\frac{1}{3}\, (24\gamma+5\delta^2)\, s^3$, for generic values of $(\gamma, \delta)$. 
\par The Weierstrass equations (\ref{Weierstrass general in 3.1.1}) and (\ref{Weierstrass the other in 3.1.1}) have the Mordell--Weil rank 0 for generic values of the parameters. For special values of the parameters, the Mordell--Weil rank is enhanced to 1, or the $ADE$ rank is enhanced to 18. K3 surfaces (\ref{Weierstrass general in 3.1.1}) and (\ref{Weierstrass the other in 3.1.1}) become attractive in these situations. Either of the equations (\ref{Weierstrass general in 3.1.1}) and (\ref{Weierstrass the other in 3.1.1}) has two parameters, since there is a parameter which determines the locations of the type $I_1$ fibers, in addition to the parameter which determines the complex structure.

\subsubsection{Enhancement to models with $U(1)$}
\label{subsec3.1.2}
\par We particularly consider the case $\beta=0$ of the Weierstrass equation (\ref{Weierstrass general in 3.1.1}). The Weierstrass equation (\ref{Weierstrass general in 3.1.1}), in this case, reduces to the following Weierstrass form:
\begin{equation}
\label{no3 Weierstrass in 3.1.2}
y^2=x^3+t^2s^3\, (\alpha\, t^3-3s^3)\, x+t^3s^6\, (\alpha\, t^3-2s^3),
\end{equation}
with the discriminant 
\begin{equation}
\label{discno3 in 3.1.2}
\Delta \sim t^{12}s^9\, (4\alpha\, t^3-9s^3).
\end{equation}
The parameter $\alpha$ parameterizes the locations of 3 type $I_1$ fibers. The complex structure of the attractive K3 surface (\ref{no3 Weierstrass in 3.1.2}) does not depend on the parameter $\alpha$, however, we must require that 
\begin{equation}
\alpha\ne 0. 
\end{equation}
After some computation, we find that the Weierstrass form (\ref{no3 Weierstrass in 3.1.2}) admits a section 
\begin{equation}
\label{torsional section in 3.1.2}
[X:Y:Z]=[-ts^3 : 0 : 1],
\end{equation}
in addition to a constant zero section
\begin{equation}
[X:Y:Z]=[0:1:0].
\end{equation}
Section (\ref{torsional section in 3.1.2}) is a torsion section that generates a $\Z_2$ group. 
\par The general form of the Weierstrass equation of an elliptic fibration with the Mordell--Weil rank 1 is determined in \cite{MorrisonPark}. By comparing the coefficients of the Weierstrass form (\ref{no3 Weierstrass in 3.1.2}) with the general form obtained in \cite{MorrisonPark}, we deduce that the Weierstrass form (\ref{no3 Weierstrass in 3.1.2}) admits another section, which generates the group $\Z$:
\begin{equation}
[X:Y:Z]=[2ts^3 : \sqrt{3\alpha}\, t^3s^3 : 1].
\end{equation}
Thus, we conclude that the Weierstrass equation (\ref{no3 Weierstrass in 3.1.2}) has the Mordell--Weil rank 1. 
\par The torsion parts of the Mordell--Weil groups of the elliptic K3 surfaces with a section were determined in \cite{Shimada}. According to Table 1 in \cite{Shimada}, the torsion part of the Mordell--Weil group of a K3 surface with $ADE$ type $E_7 D_{10}$ is either 0 or $\Z_2$ (No.2421 in Table 1 in \cite{Shimada}). Thus, we conclude that the $\Z_2$ group generated by the torsion section (\ref{torsional section in 3.1.2}) gives the whole torsion part of the Mordell--Weil group of the K3 surface (\ref{no3 Weierstrass in 3.1.2}). Therefore, we deduce that the Mordell--Weil group of the K3 surface (\ref{no3 Weierstrass in 3.1.2}) is isomorphic to $\Z\oplus\Z_2$.
\par Thus, we conclude that the Weierstrass equation (\ref{no3 Weierstrass in 3.1.2}) describes an attractive K3 surface with the $ADE$ type $E_7 D_{10}$ with the Mordell--Weil group $\Z\oplus\Z_2$. Comparing the equation (\ref{no3 Weierstrass in 3.1.2}) with the general Weierstrass equation of the fibration No.3 given in \cite{Ut}, we deduce that the equation (\ref{no3 Weierstrass in 3.1.2}) gives the fibration No.3 in Table \ref{tabfibrationX3} of the attractive K3 surface $X_3$. 
\par Therefore, we find that the Mordell--Weil rank is enhanced to 1 at the point $\beta=0$ in moduli of K3 surfaces with the $ADE$ type $E_7 D_{10}$ (\ref{Weierstrass general in 3.1.1}). The K3 surface becomes attractive at this point. F-theory compactification on the K3 elliptic fibration (\ref{no3 Weierstrass in 3.1.2}) times a K3 surface has a $U(1)$ gauge field.

\vspace{5mm}

\par Using the Kneser-Nishiyama method, we find that the attractive K3 surfaces that admit an elliptic fibration with a section with $ADE$ type $E_7 D_{10}$ include: attractive K3 surfaces, the intersection matrices of the transcendental lattices of which are $\begin{pmatrix}
4 & 0 \\
0 & 2 \\
\end{pmatrix}$, $\begin{pmatrix}
6 & 0 \\
0 & 2 \\
\end{pmatrix}$, $\begin{pmatrix}
4 & 1 \\
1 & 2 \\
\end{pmatrix}$, $\begin{pmatrix}
8 & 1 \\
1 & 2 \\
\end{pmatrix}$, and $\begin{pmatrix}
12 & 0 \\
0 & 2 \\
\end{pmatrix}$ \footnote{Elliptic fibrations with a global section, and their Weierstrass forms, of an attractive K3 surface whose transcendental lattice has the intersection matrix $\begin{pmatrix}
4 & 0 \\
0 & 2 \\
\end{pmatrix}$ are classified in \cite{BLe}. Elliptic fibrations with a global section of an attractive K3 surface whose transcendental lattice has the intersection matrix $\begin{pmatrix}
6 & 0 \\
0 & 2 \\
\end{pmatrix}$ are studied in \cite{BGHLMSW}.}, in addition to the attractive K3 surface $X_3$. These attractive K3 surfaces correspond to special points in the moduli of K3 surfaces (\ref{Weierstrass general in 3.1.1}) and (\ref{Weierstrass the other in 3.1.1}) at which the Mordell--Weil ranks are enhanced. We particularly discuss the attractive K3 surface whose transcendental lattice has the intersection matrix $\begin{pmatrix}
4 & 0 \\
0 & 2 \\
\end{pmatrix}$.
\par The general Weierstrass equation of the elliptic fibration with $ADE$ type $E_7 D_{10}$ of the attractive K3 surface whose transcendental lattice has the intersection matrix $\begin{pmatrix}
4 & 0 \\
0 & 2 \\
\end{pmatrix}$ is given in \cite{BLe} as:
\begin{equation}
\label{general Weierstrass 402 in 3.1.2}
y^2=x^3-(5t^2+t)\, x^2-t^5\, x.
\end{equation}
The general Weierstrass equation (\ref{general Weierstrass 402 in 3.1.2}) transforms into the following Weierstrass equation:
\begin{equation}
\label{Weierstrass 402 of type 1 in 3.1.2}
y^2=x^3-t^2s^3\, [\, t^3+\frac{1}{3}s(5t+s)^2 \, ]\, x-\frac{1}{27}t^3(5t+s)\, (9t^3+50t^2s+20ts^2+2s^3)\, s^5.
\end{equation}
Under the following rescaling
\begin{eqnarray}
\label{rescaling root 3 in 3.1.2}
x & \rightarrow & \frac{x}{(\sqrt{3})^2} \\ \nonumber
y & \rightarrow & \frac{y}{(\sqrt{3})^3},
\end{eqnarray}
the coefficients $f,g$ of the Weierstrass form $y^2=x^3+f\, x+g$ transform as
\begin{eqnarray}
f & \rightarrow & 3^2\, f \\ \nonumber
g & \rightarrow & 3^3\, g.
\end{eqnarray}
Thus, under the rescaling (\ref{rescaling root 3 in 3.1.2}), the Weierstrass form (\ref{Weierstrass 402 of type 1 in 3.1.2}) further transforms into another Weierstrass form as:
\begin{equation}
\label{Weierstrass 402 of type 2 in 3.1.2}
y^2=x^3-t^2s^3\, [\, 9t^3+3s(5t+s)^2 \, ]\, x-t^3(5t+s)\, (9t^3+50t^2s+20ts^2+2s^3)\, s^5.
\end{equation}
Weierstrass equation (\ref{Weierstrass 402 of type 2 in 3.1.2}) is a member of the family of the Weierstrass equations with $ADE$ type $E_7 D_{10}$ (\ref{Weierstrass general in 3.1.1}), with  
\begin{eqnarray}
\alpha & = & -9 \\ \nonumber
\beta & = & -30. 
\end{eqnarray}
Thus, we conclude that the elliptic fibration with the $ADE$ type $E_7 D_{10}$ of the attractive K3 surface whose transcendental lattice has the intersection matrix $\begin{pmatrix}
4 & 0 \\
0 & 2 \\
\end{pmatrix}$ is described by the equation (\ref{Weierstrass general in 3.1.1}) with $(\alpha, \beta)=(-9, -30)$. 
\par The Mordell--Weil group of the fibration (\ref{general Weierstrass 402 in 3.1.2}) is isomorphic to $\Z\oplus \Z_2$ \cite{BLe}. A $U(1)$ gauge field arises in F-theory compactification on the K3 elliptic fibration (\ref{Weierstrass 402 of type 2 in 3.1.2}) times a K3 surface. 

\subsubsection{Enhancement to rank 18}
\label{subsec3.1.3}
The $ADE$ rank of a K3 surface in the moduli (\ref{Weierstrass general in 3.1.1}) and (\ref{Weierstrass the other in 3.1.1}) is enhanced to 18 at special points. The K3 surface becomes attractive at these points. The Mordell--Weil rank remains to be 0 in these situations. We discuss some such points. 
\par When $\delta=0$ in the moduli (\ref{Weierstrass the other in 3.1.1}), the Weierstrass equation (\ref{Weierstrass the other in 3.1.1}) becomes:
\begin{equation}
\label{E8D10 in 3.1.3}
y^2 = x^3 + t^2s^4\, ( \gamma \, t^2 -3s^2 )\, x +t^3s^5\, ( -\frac{\gamma^2}{12}\, t^4+\gamma\, t^2s^2-2s^4 ).
\end{equation}
The discriminant of the Weierstrass equation (\ref{E8D10 in 3.1.3}) is
\begin{equation}
\Delta \sim t^{12}s^{10}\, (3\gamma t^2-8s^2).
\end{equation}
The $ADE$ type of the Weierstrass equation (\ref{E8D10 in 3.1.3}) is $E_8 D_{10}$; a type $II^*$ fiber is at $[t:s]=[1:0]$, and a type $I^*_6$ fiber is at $[t:s]=[0:1]$. The K3 surface (\ref{E8D10 in 3.1.3}) is attractive with the Mordell--Weil rank 0. We find from Table 2 in \cite{SZ} that the attractive K3 surface (\ref{E8D10 in 3.1.3}) is in fact the attractive K3 surface with the discriminant 4, $X_4$ (No.320 in Table 2 in \cite{SZ}). Therefore, we deduce that the Weierstrass equation (\ref{E8D10 in 3.1.3}) gives the fibration No.2 in Table \ref{tabfibrationX4}. (In the equation (\ref{E8D10 in 3.1.3}), we require that $\gamma\ne 0$.)

\vspace{5mm}

\par We find from equations (\ref{Weierstrass the other in 3.1.1}) and (\ref{disc the other in 3.1.1}) that when the parameters of the Weierstrass equation (\ref{Weierstrass the other in 3.1.1}) satisfy
\begin{equation}
\label{condition 3.1.3}
24\gamma+5\delta^2=0,
\end{equation}
the $ADE$ type is enhanced to $E_7 D_{11}$. (We exclude the case $\gamma=\delta=0$.) We deduce from Table 2 in \cite{SZ} that, in this situation, the K3 surface (\ref{Weierstrass the other in 3.1.1}) becomes the attractive K3 surface whose transcendental lattice has the intersection matrix $\begin{pmatrix}
4 & 0 \\
0 & 2 \\
\end{pmatrix}$ (No.291 in Table 2 in \cite{SZ}). 
Therefore, we conclude that the K3 surface becomes the attractive K3 surface whose transcendental lattice has the intersection matrix $\begin{pmatrix}
4 & 0 \\
0 & 2 \\
\end{pmatrix}$ in the moduli (\ref{Weierstrass the other in 3.1.1}) when the parameters satisfy the condition (\ref{condition 3.1.3}). The Mordell--Weil group of this K3 elliptic fibration is isomorphic to 0 \cite{SZ}.

\vspace{5mm}

\par We find from the equations (\ref{Weierstrass general in 3.1.1}) and (\ref{disc E7D10 general in 3.1.1}) that when $(\alpha, \beta)=(1, 6\sqrt[3]{3})$, the $ADE$ type is enhanced to $E_7 D_{10} A_1$. The K3 surface (\ref{Weierstrass general in 3.1.1}) becomes attractive in this situation. We deduce from Table 2 in \cite{SZ} that this attractive K3 surface is the attractive K3 surface $X_4$. (No.290 in Table 2 in \cite{SZ}) Therefore, we find that a K3 surface becomes the surface $X_4$ at $(\alpha, \beta)=(1, 6\sqrt[3]{3})$ in the moduli (\ref{Weierstrass general in 3.1.1}). The Weierstrass equation (\ref{Weierstrass general in 3.1.1}) with $(\alpha, \beta)=(1, 6\sqrt[3]{3})$ describes the fibration No.5 in Table \ref{tabfibrationX4}.

\subsection{Moduli of K3 surfaces with $ADE$ type $A_{17}$}
\label{subsec3.2}

\subsubsection{Weierstrass equation of K3 surfaces with $ADE$ type $A_{17}$}
\label{subsec3.2.1}
\par The method to deduce the Weierstrass equation of a K3 surface from the Weierstrass equation of a rational elliptic surface, by gluing two isomorphic rational elliptic surfaces, is discussed in \cite{KRES}. Using this method, we construct a family of Weierstrass equations that describes K3 surfaces with $ADE$ type $A_{17}$. 
\par Gluing two identical rational elliptic surfaces along smooth fibers, we generically obtain a K3 surface, the singular fibers of which are twice the singular fibers of the rational elliptic surface. This corresponds to the quadratic base change of the rational elliptic surface \cite{KRES}. 
\par We consider the extremal rational elliptic surface with four singular fibers, of types $I_9$, $I_1$, $I_1$, and $I_1$. Following the notation used in \cite{KRES}, we denote this extremal rational elliptic surface by $X_{[9, 1, 1, 1]}$. The Weierstrass form of the extremal rational elliptic surface $X_{[9, 1, 1, 1]}$ is given as follows \cite{MP}:
\begin{equation}
\label{RES 9111 in 3.2.1}
y^2=x^3-3t\, (t^3+24s^3)\, x+2(t^6+36t^3s^3+216s^6).
\end{equation}
We consider the gluing of two copies of the extremal rational elliptic surface $X_{[9, 1, 1, 1]}$.
\par The limit of the quadratic base change of rational elliptic surfaces, at which singular fibers of the same type collide, is also discussed in \cite{KRES}. We consider the limit of the gluing of two identical extremal rational elliptic surfaces at which two $I_9$ fibers collide; they are enhanced to an $I_{18}$ fiber in this limit. This limit is given by the following base change:
\begin{eqnarray}
\label{base change no4 in 3.2.1}
t & \rightarrow & t^2+\alpha\,  s^2 \\ \nonumber
s & \rightarrow & s^2. 
\end{eqnarray}
In transformation (\ref{base change no4 in 3.2.1}), $\alpha$ is a parameter. The transformation of $t$, in general, can take the form $t \rightarrow t^2+\alpha\, s^2+\gamma\, ts$. However, by completing the square in $t$, we find that $\gamma$ is a redundant parameter. The K3 surface as a result of the base change (\ref{base change no4 in 3.2.1}) has the following Weierstrass form:
\begin{equation}
\label{no4 Weierstrass in 3.2.1}
\begin{split}
y^2= & x^3-3(t^2+\alpha\, s^2)\, [\, (t^2+\alpha\, s^2)^3+24s^6\, ]\, x \\
 & +2\, [\, (t^2+\alpha\, s^2)^6+36(t^2+\alpha\, s^2)^3s^6+216 s^{12}\, ].
 \end{split}
\end{equation}
The discriminant of the Weierstrass equation (\ref{no4 Weierstrass in 3.2.1}) is given by 
\begin{equation}
\label{disc no4 in 3.2.1}
\Delta \sim s^{18}\, [\, (t^2+\alpha\, s^2)^3+27 s^6\, ].
\end{equation}
From equations (\ref{no4 Weierstrass in 3.2.1}) and (\ref{disc no4 in 3.2.1}), we confirm that the K3 surface (\ref{no4 Weierstrass in 3.2.1}) has 1 type $I_{18}$ fiber at $[t:s]=[1:0]$, and 6 type $I_1$ fibers at the roots of $ (t^2+\alpha\, s^2)^3+27 s^6$, for generic values of $\alpha$. 
\par For generic values of $\alpha$, the Mordell--Weil group of the Weierstrass equations (\ref{no4 Weierstrass in 3.2.1}) has rank 0; for special values of $\alpha$, the Mordell--Weil rank is enhanced, becoming 1, or the $ADE$ rank is enhanced to 18. For these special values of $\alpha$, the K3 surface (\ref{no4 Weierstrass in 3.2.1}) becomes attractive.

\subsubsection{Enhancement to models with $U(1)$}
\label{subsec3.2.2}
\par We particularly consider the case $\alpha=0$. For this case, the equation (\ref{no4 Weierstrass in 3.2.1}) becomes the following Weierstrass form:
\begin{equation}
\label{no4 Weierstrass specific in 3.2.2}
y^2=x^3-3t^2\, (t^6+24 s^6)\, x+2\,(t^{12}+36\, t^6s^6+216\, s^{12}).
\end{equation}
It can be confirmed that the Weierstrass equation (\ref{no4 Weierstrass specific in 3.2.2}) has the following sections:
\begin{eqnarray}
\label{torsions in 3.2.2}
[X:Y:Z]& = & [t^4: 12\sqrt{3}\, s^6:1], \\ \nonumber
 & & [t^4: -12\sqrt{3}\, s^6:1].
\end{eqnarray}
Sections (\ref{torsions in 3.2.2}) are torsion sections, and they generate the group $\Z_3$. The torsion part of the Mordell--Weil group of an elliptic K3 surface with $ADE$ type $A_{17}$ is either 0 or $\Z_3$ \cite{Shimada}. (See No.2787 in Table 1 in \cite{Shimada}.) Therefore, we conclude that $\Z_3$ group generated by torsion sections (\ref{torsions in 3.2.2}) gives the whole torsion part of the Mordell--Weil group of the K3 surface (\ref{no4 Weierstrass specific in 3.2.2}). 
\par Some computation shows that, in addition to the torsion sections (\ref{torsions in 3.2.2}), the K3 surface (\ref{no4 Weierstrass specific in 3.2.2}) admits another section:
\begin{equation}
\label{new section in 3.2.2}
[X:Y:Z]=[t^4+12\, ts^3: 12\sqrt{3}\,(t^3s^3+s^6): 1].
\end{equation}
Section (\ref{new section in 3.2.2}) does not belong to the torsion part $\Z_3$. This shows that the Mordell--Weil group of the K3 surface (\ref{no4 Weierstrass specific in 3.2.2}) has a free part. Thus, we conclude that the K3 surface (\ref{no4 Weierstrass specific in 3.2.2}) has the Mordell--Weil rank 1. 
\par The Weierstrass equation (\ref{no4 Weierstrass specific in 3.2.2}) gives an attractive K3 surface with the $ADE$ type $A_{17}$ with the Mordell--Weil group isomorphic to $\Z\oplus\Z_3$.
\par We find that the equation (\ref{no4 Weierstrass specific in 3.2.2}) describes the fibration No.4 of the attractive K3 surface $X_3$ in Table \ref{tabfibrationX3} \cite{Ut}. F-theory compactification on the K3 elliptic fibration (\ref{no4 Weierstrass specific in 3.2.2}) times a K3 surface has a $U(1)$ gauge field.

\vspace{5mm}

\par We find via the Kneser-Nishiyama method that the attractive K3 surfaces that admit an elliptic fibration with a section with the $ADE$ type $A_{17}$ include: attractive K3 surfaces, the intersection matrices of the transcendental lattices of which are $\begin{pmatrix}
4 & 0 \\
0 & 2 \\
\end{pmatrix}$, $\begin{pmatrix}
6 & 0 \\
0 & 2 \\
\end{pmatrix}$, $\begin{pmatrix}
4 & 1 \\
1 & 2 \\
\end{pmatrix}$, $\begin{pmatrix}
8 & 1 \\
1 & 2 \\
\end{pmatrix}$, and $\begin{pmatrix}
12 & 0 \\
0 & 2 \\
\end{pmatrix}$, in addition to the attractive K3 surface $X_3$. 
\par We show that the Weierstrass equation (\ref{no4 Weierstrass in 3.2.1}) with 
\begin{equation}
\alpha=5
\end{equation}
gives the attractive K3 surface whose transcendental lattice has the intersection matrix $\begin{pmatrix}
4 & 0 \\
0 & 2 \\
\end{pmatrix}$. 
The general Weierstrass form of elliptic fibration with the $ADE$ type $A_{17}$ of the attractive K3 surface whose transcendental lattice has the intersection matrix $\begin{pmatrix}
4 & 0 \\
0 & 2 \\
\end{pmatrix}$ is given in \cite{BLe} as:
\begin{equation}
\label{general Weierstrass 402 in 3.2.2}
y^2+(t^2+5)\, xy+y=x^3.
\end{equation}
The general Weierstrass form (\ref{general Weierstrass 402 in 3.2.2}) transforms into the following Weierstrass form:
\begin{equation}
\label{Weierstrass 402 type 1 in 3.2.2}
\begin{split}
y^2 = & x^3 - \frac{1}{2}(t^2+5s^2)\, [\, \frac{1}{24}(t^2+5s^2)^3-s^6 \, ]\, x \\
& +\frac{2}{12^3}(t^2+5s^2)^6 - \frac{1}{24}(t^2+5s^2)^3\, s^6+\frac{s^{12}}{4}. 
\end{split}
\end{equation}
Under the following rescaling
\begin{eqnarray}
x & \rightarrow & \frac{x}{(\sqrt{12})^2} \\ \nonumber
y & \rightarrow & \frac{y}{(\sqrt{12})^3},
\end{eqnarray}
the Weierstrass form (\ref{Weierstrass 402 type 1 in 3.2.2}) further transforms into another Weierstrass form as
\begin{equation}
\label{Weierstrass 402 type 2 in 3.2.2}
\begin{split}
y^2 = & x^3 -3(t^2+5s^2)\, [\, (t^2+5s^2)^3-24\, s^6 \, ]\, x \\
& +2\, [\, (t^2+5s^2)^6 - 36(t^2+5s^2)^3\, s^6+216\, s^{12} \, ]. 
\end{split}
\end{equation}
The Weierstrass equation (\ref{Weierstrass 402 type 2 in 3.2.2}) corresponds to the following base change of the rational elliptic surface $X_{[9, 1, 1, 1]}$ (\ref{RES 9111 in 3.2.1}):
\begin{eqnarray}
\label{base change 9111 in 3.2.2}
t & \rightarrow & -(t^2+5\,  s^2) \\ \nonumber
s & \rightarrow & s^2. 
\end{eqnarray}
When we replace $t$ with $it$ and $\alpha$ with $-\alpha$, $-(t^2+\alpha s^2)$ is replaced with $t^2+\alpha s^2$. Therefore, the minus sign of $-(t^2+5 s^2)$ in the base change (\ref{base change 9111 in 3.2.2}) may be replaced by a positive sign. Thus, we deduce that the K3 surface (\ref{no4 Weierstrass in 3.2.1}) becomes an attractive K3 surface whose transcendental lattice has the intersection matrix $\begin{pmatrix}
4 & 0 \\
0 & 2 \\
\end{pmatrix}$ when $\alpha$ takes the value 5. 
\par The Mordell--Weil group of the elliptic fibration (\ref{general Weierstrass 402 in 3.2.2}) is isomorphic to $\Z\oplus \Z_3$ \cite{BLe}. Thus, F-theory compactification on the K3 elliptic fibration (\ref{Weierstrass 402 type 2 in 3.2.2}) times a K3 surface has a $U(1)$ gauge field. 

\subsubsection{Enhancement to rank 18}
\label{subsec3.2.3}
\par We find from the equations (\ref{no4 Weierstrass in 3.2.1}) and (\ref{disc no4 in 3.2.1}) that the $ADE$ type is enhanced to $A_{17} A_1$ when $\alpha=-3$. The type $I_{18}$ fiber is at $[t:s]=[1:0]$, and the type $I_2$ fiber is at $[t:s]=[0:1]$. The K3 surface (\ref{no4 Weierstrass in 3.2.1}) becomes attractive in this situation. According to Table 1 in \cite{Shimada}, the Mordell--Weil group of an attractive K3 surface with $ADE$ type $A_{17} A_1$ is isomorphic to either 0 or $\Z_3$. K3 surface (\ref{no4 Weierstrass in 3.2.1}) always has torsion sections:
\begin{eqnarray}
\label{torsions in 3.2.3}
[X:Y:Z]& = & [(t^2+\alpha \, s^2)^2: 12\sqrt{3}\, s^6:1], \\ \nonumber
 & & [(t^2+\alpha \, s^2)^2: -12\sqrt{3}\, s^6:1].
\end{eqnarray}
The torsion sections (\ref{torsions in 3.2.3}) generate the group $\Z_3$. Thus, we conclude that the Mordell--Weil group of the attractive K3 surface (\ref{no4 Weierstrass in 3.2.1}) with $\alpha=-3$ is isomorphic to $\Z_3$. Therefore, the K3 surface (\ref{no4 Weierstrass in 3.2.1}) with $\alpha=-3$ has the $ADE$ type $A_{17} A_1$ with the Mordell--Weil group $\Z_3$. We find from Table 2 in \cite{SZ} that the K3 surface (\ref{no4 Weierstrass in 3.2.1}) with $\alpha=-3$ is the attractive K3 surface $X_4$. (No.111 in Table 2 in \cite{SZ}) Therefore, K3 surface in the moduli (\ref{no4 Weierstrass in 3.2.1}) becomes the attractive K3 surface $X_4$ at $\alpha=-3$. The Weierstrass equation (\ref{no4 Weierstrass in 3.2.1}) with $\alpha=-3$ describes the fibration No.6 in Table \ref{tabfibrationX4}.

\section{Gauge symmetries in F-theory compactifications}
\label{sec 4}

\subsection{Gauge symmetries in F-theory models on special points in the moduli of K3 surfaces with $ADE$ types $E_7 D_{10}$ and $A_{17}$}
\label{subsec4.1}
\par In section \ref{sec 3}, we discussed special points in the moduli of K3 surfaces with $ADE$ types $E_7 D_{11}$ and $A_{17}$, at which the Picard number is enhanced to 20 and the K3 surfaces become attractive. We consider F-theory compactifications on these attractive K3 surfaces times a K3 surface. The $ADE$ types and Mordell--Weil groups of these attractive K3 surfaces were determined in section \ref{sec 3}. The global structures of the gauge groups in F-theory compactifications follow from these results. We summarize the structures of the gauge symmetries in Tables \ref{tabgaugeE7D10} and \ref{tabgaugeA17}.

\begingroup
\renewcommand{\arraystretch}{1.5}
\begin{table}[htb]
\begin{center}
  \begin{tabular}{|c|c|c|c|c|} \hline
$
\begin{array}{c}
\mbox{Weierstrass}\\
\mbox{equation} 
\end{array}
$  & Complex Str. & Mordell-Weil group & Gauge group \\ \hline
equation (\ref{no3 Weierstrass in 3.1.2}) & $X_3$ & $\Z\oplus\Z_2$ & $E_7 \times SO(20) / \Z_2 \times U(1)$ \\ \hline
$
\begin{array}{c}
\mbox{equation (\ref{Weierstrass general in 3.1.1})}\\
\mbox{with $(\alpha, \beta)=(-9, -30)$} 
\end{array}
$ & $\begin{pmatrix}
4 & 0 \\
0 & 2 \\
\end{pmatrix}$ & $\Z\oplus\Z_2$ & $E_7\times SO(20) / \Z_2  \times U(1)$ \\ \hline
$
\begin{array}{c}
\mbox{equation (\ref{Weierstrass the other in 3.1.1})}\\
\mbox{with $24\gamma+5\delta^2=0$} 
\end{array}
$ & $\begin{pmatrix}
4 & 0 \\
0 & 2 \\
\end{pmatrix}$ & 0 & $E_7 \times SO(22)$ \\ \hline
equation (\ref{E8D10 in 3.1.3}) & $X_4$ & 0 & $E_8\times SO(20)$ \\ \hline
$
\begin{array}{c}
\mbox{equation (\ref{Weierstrass general in 3.1.1})}\\
\mbox{with $(\alpha, \beta)=(1, 6\sqrt[3]{3})$} 
\end{array}
$ & $X_4$ & $\Z_2$ & $E_7 \times SO(20) \times SU(2) / \Z_2$ \\ \hline
\end{tabular}
\caption{\label{tabgaugeE7D10}Some attractive K3 surfaces that appear at special enhanced points in the moduli of K3 surfaces with $ADE$ type $E_7 D_{10}$, and the structures of the gauge symmetries in F-theory compactifications on these attractive K3 surfaces times a K3 surface.}
\end{center}
\end{table}  
\endgroup

\begingroup
\renewcommand{\arraystretch}{1.5}
\begin{table}[htb]
\begin{center}
  \begin{tabular}{|c|c|c|c|c|} \hline
$
\begin{array}{c}
\mbox{Weierstrass}\\
\mbox{equation} 
\end{array}
$  & Complex Str. & Mordell-Weil group & Gauge group \\ \hline
equation (\ref{no4 Weierstrass specific in 3.2.2}) & $X_3$ & $\Z\oplus\Z_3$ & $SU(18) / \Z_3 \times U(1)$ \\ \hline
$
\begin{array}{c}
\mbox{equation (\ref{no4 Weierstrass in 3.2.1})}\\
\mbox{with $\alpha=5$} 
\end{array}
$ & $\begin{pmatrix}
4 & 0 \\
0 & 2 \\
\end{pmatrix}$ & $\Z\oplus\Z_3$ & $SU(18) / \Z_3  \times U(1)$ \\ \hline
$
\begin{array}{c}
\mbox{equation (\ref{no4 Weierstrass in 3.2.1})}\\
\mbox{with $\alpha=-3$} 
\end{array}
$ & $X_4$ & $\Z_3$ & $SU(18) \times SU(2) / \Z_3$ \\ \hline
\end{tabular}
\caption{\label{tabgaugeA17}Some attractive K3 surfaces that appear at special enhanced points in the moduli of K3 surfaces with $ADE$ type $A_{17}$, and the structures of the gauge symmetries in F-theory compactifications on these attractive K3 surfaces times a K3 surface.}
\end{center}
\end{table}  
\endgroup

\subsection{Cancellation of anomaly}
\label{subsec4.2}
\par In section \ref{subsec4.1}, we considered F-theory compactified on spaces which are built as the direct products of K3 surfaces. Such compactifications give a four-dimensional theory with $N=2$ supersymmetry. As we stated in section \ref{subsec3.1.1}, it can be deduced from the anomaly cancellation condition that 7-branes are wrapped on K3 surfaces, and there are 24 7-branes in F-theory on K3 $\times$ K3. We determined the configurations of the singular fibers of the K3 surfaces with the $ADE$ type $E_7 D_{10}$ from the anomaly cancellation condition in section \ref{subsec3.1.1}. We confirm that the anomaly cancellation condition is satisfied for F-theory compactifications on K3 surfaces with the $ADE$ type $A_{17}$ (\ref{no4 Weierstrass in 3.2.1}) times a K3. We determined in section \ref{subsec3.2.1} that the K3 surface with the $ADE$ type $A_{17}$ (\ref{no4 Weierstrass in 3.2.1}) has 1 type $I_{18}$ and 6 type $I_1$ fibers. According to Table \ref{tab7-branesnumber}, the number of associated 7-branes is 24. This satisfies the anomaly cancellation condition. As we saw in section \ref{sec 3}, the K3 surface becomes attractive at special points in the moduli of K3 surfaces with the $ADE$ types $E_7 D_{10}$ and $A_{17}$. An argument similar to that stated above shows that the anomaly cancellation condition is satisfied at these points. 
\par We also discuss the cancellation of the tadpole with 4-form flux turned on \cite{SVW} at special points at which K3 surfaces become attractive. By including 4-form flux \cite{BB, SVW, W, DRS}, F-theory compactifications on the direct products of K3 surfaces become four-dimensional theory with $N=1$ supersymmetry. We saw in section \ref{sec 3} that the attractive K3 surfaces $X_3$, $X_4$ and the attractive K3 surfaces, the intersection matrices of the transcendental lattices of which are $\begin{pmatrix}
4 & 0 \\
0 & 2 \\
\end{pmatrix}$, $\begin{pmatrix}
6 & 0 \\
0 & 2 \\
\end{pmatrix}$, $\begin{pmatrix}
4 & 1 \\
1 & 2 \\
\end{pmatrix}$, $\begin{pmatrix}
8 & 1 \\
1 & 2 \\
\end{pmatrix}$, and $\begin{pmatrix}
12 & 0 \\
0 & 2 \\
\end{pmatrix}$, appear in the moduli of K3 surfaces with the $ADE$ types $E_7 D_{10}$ and $A_{17}$. We deduce from Table 1 in \cite{AK} and Table 2 in \cite{BKW} that for these attractive K3 surfaces, when they are paired with some attractive K3 surfaces with appropriate complex structures, the tadpole can be cancelled in F-theory compactifications on these attractive K3 surfaces times such appropriately chosen attractive K3 surfaces. Discussions of the cancellation of the tadpole in F-theory flux compactifications on K3 $\times$ K3 can be found in \cite{K, K2}.

\section{Conclusion}
\label{sec 5}
We constructed the moduli of K3 surfaces with the $ADE$ types $E_7 D_{10}$ and $A_{17}$. The Picard number is enhanced to 20 at special points in these moduli; the K3 surfaces become attractive at such points. At these points, either the Mordell--Weil rank increases, or the non-Abelian gauge symmetry on the 7-branes in F-theory compactification is enhanced. We determined some of the special points at which K3 surfaces become attractive in the moduli of K3 surfaces with the $ADE$ types $E_7 D_{10}$ and $A_{17}$. We also studied the gauge groups in F-theory compactifications on the attractive K3 surfaces, which correspond to the special points in the moduli times a K3 surface. A $U(1)$ gauge symmetry arises in F-theory compactifications at the points in the moduli where the Mordell--Weil rank is enhanced.

\section*{Acknowledgments}
We would like to thank Taro Tani for discussions. We are also grateful to the referee for improving this manuscript. YK is partially supported by Grant-in-Aid for Scientific Research {\#}16K05337 from the Ministry of Education, Culture, Sports, Science and Technology of Japan. SM is supported by Grant-in-Aid for Scientific Research {\#}16K05337 from the Ministry of Education, Culture, Sports, Science and Technology of Japan.

\end{document}